# Superconductivity across Lifshitz transition and anomalous insulating state in surface K-dosed (Li$_{0.8}$Fe$_{0.2}$OH)FeSe

M. Q. Ren[1], Y. J. Yan[1], X. H. Niu[1], R. Tao[1], D. Hu[1], R. Peng[1], B. P. Xie[1,2], J. Zhao[1,2], T. Zhang[1,2*], D. L. Feng[1,2*]

[1]State Key Laboratory of Surface Physics, Department of Physics, and Advanced Materials Laboratory, Fudan University, Shanghai 200433, China
[2]Collaborative Innovation Center of Advanced Microstructures, Nanjing, 210093, China
*Email: tzhang18@fudan.edu.cn, dlfeng@fudan.edu.cn

**In the iron-based superconductors, understanding the relation between superconductivity and electronic structure upon doping is crucial for exploring the pairing mechanism. Recently it was found that in iron selenide (FeSe), enhanced superconductivity ($T_c$ over 40K) can be achieved via electron doping, with the Fermi surface only comprising M-centered electron pockets. Here by utilizing surface potassium dosing, scanning tunneling microscopy/spectroscopy (STM/STS) and angle-resolved photoemission spectroscopy (ARPES), we studied the electronic structure and superconductivity of (Li$_{0.8}$Fe$_{0.2}$OH)FeSe in the deep electron-doped regime. We find that a Γ-centered electron band, which originally lies above the Fermi level ($E_F$), can be continuously tuned to cross $E_F$ and contribute a new electron pocket at Γ. When this Lifshitz transition occurs, the superconductivity in the M-centered electron pocket is slightly suppressed; while a possible superconducting gap with small size (up to ~5 meV) and a dome-like doping dependence is observed on the new Γ electron pocket. Upon further K dosing, the system eventually evolves into an insulating state. Our findings provide new clues to understand superconductivity versus Fermi surface topology and the correlation effect in FeSe-based superconductors.**

In high-$T_c$ iron-based superconductors, carrier doping is one of the principal routes to induce superconductivity. Many factors such as the density of states (DOS), Fermi surface topology and nesting condition, and correlation strength may vary significantly with carrier concentration. Detailed knowledge of the electronic structure vs. doping is critical for understanding the pairing mechanism. Recently, it was found that through heavy electron doping, the $T_c$ of FeSe can be enhanced from the bulk value of 8K to over 40K. The doping can be achieved via interlayer intercalation [A$_x$Fe$_{2-y}$Se$_2$ (A=K, Rb …)(1,2), (Li,NH$_3$)FeSe (3), (Li$_{1-x}$Fe$_x$OH)FeSe (4)], interface charge transfer (FeSe/SrTiO$_3$) (5), surface K-dosing (6), and ionic-liquid gating (7~9). ARPES studies show that $T_c$ enhancement in these systems is universally accompanied by a vanishing of the Γ hole pockets, and that the superconducting gap on the M electron pockets is nodeless (10~14). Meanwhile, STM studies suggest that the pairing symmetries of single-layer FeSe/SrTiO$_3$ and (Li$_{0.8}$Fe$_{0.2}$OH)FeSe are plain $s$-wave (15,16), which differs from the $s_\pm$-wave of bulk FeSe and FeTe$_x$Se$_{1-x}$ (17,18), and double-dome-like superconductivity is observed in FeSe films upon K-dosing (19). These results indicate that the high-$T_c$ phase in heavily electron-doped FeSe may be quite different from that in undoped FeSe, with changes in Fermi surface topology likely playing a crucial role.

Despite the $T_c$ enhancement, the detailed phase diagram of electron-doped FeSe, particularly in the region beyond "optimal" doping, is still not fully understood. Recent ARPES results show that after FeSe films enter the high-$T_c$ phase *via* surface K-dosing, the electron correlation anomalously increases upon further doping and eventually an insulating phase emerges (20). This indicates remarkable complexity and new physics in the "overdoped" region. In this work, by using low-temperature STM and ARPES, we studied the detailed evolution of the superconductivity and electronic structure of $(Li_{0.8}Fe_{0.2}OH)FeSe$ via surface potassium dosing. $(Li_{0.8}Fe_{0.2}OH)FeSe$ is already heavily electron-doped with a $T_c$ of ~40 K (4,16). Surface K dosing can further increase the doping level of the surface FeSe layer. We observe that an unoccupied, Γ-centered electron band shifts significantly to $E_F$ with increasing K coverage ($K_c$), while the double superconducting gap on M-centered electron pockets gets suppressed slightly. At certain $K_c$, the Γ-centered band crosses $E_F$, resulting in a Lifshitz transition of the Fermi surface. Shortly after the transition, a superconducting-like gap (up to 5 meV) opens at $E_F$, showing a dome-like dependence on $K_c$. This represents a new Fermi surface topology for iron-based superconductors, which has sizable electron Fermi pockets at both the Brillouin zone center and the zone corner. At even higher $K_c$, the system eventually evolves into an insulating phase, characterized by a large, asymmetric gap in excess of 50 meV. The presence of a novel Fermi surface topology, anomalous insulating phase, and the continuous tunability make $(Li_{0.8}Fe_{0.2}OH)FeSe$ a unique platform for gaining insight into the mechanism of iron-based superconductors.

**Characterization of as-cleaved FeSe surface**

$(Li_{0.8}Fe_{0.2}OH)FeSe$ single crystals with a $T_c$ of ~42K (see Fig.S1) were grown by hydrothermal reaction method (4, 21). Details of the sample preparation and STM measurement are described in Method section. There are two possible surface terminations in a cleaved sample, namely $Li_{0.8}Fe_{0.2}OH$-terminated and FeSe-terminated surfaces, as reported previously (16). Here we focus on the FeSe surface with K dosing (see Method section for details). Fig. 1A shows a topographic image of an as-cleaved FeSe surface. The square Se lattice (inset) and some dimer-shaped defects can be resolved. The dI/dV spectrum of this surface taken near $E_F$ shows a double superconducting gap (Fig. 1B). For comparison, the topographic image and STS of $Li_{0.8}Fe_{0.2}OH$ surface are shown in Fig. S2, which are distinct from the FeSe surface. The gap sizes of the FeSe surface determined from the two sets of coherence peaks are $\Delta_1 = 14.2$ meV and $\Delta_2 = 8.9$ meV, similar to previous reports (16,22). As shown by ARPES studies (13,14), these superconducting gaps are from M-centered electron pockets, while the double-peaked structure could be due to gap anisotropy (23) or band hybridization (22). The gap is found to be spatially homogeneous on FeSe surface (see Fig. S3), confirming the high quality of the sample.

Fig. 1C shows the typical dI/dV spectrum of FeSe surface on a larger energy scale (±200 meV). The tunneling conductance is relatively low near $E_F$ but increases rapidly above 70mV and below -55mV. The double superconducting gap is not observable on this scale. We note Huang *et al.* observed similar dI/dV spectra in single-layer FeSe/SrTiO$_3$ (24). They revealed that an unoccupied, Γ-centered electron band gives the steep dI/dV upturn at positive bias. This band is well reproduced in DFT calculations (24,25). The dI/dV upturn at negative bias is from the onset of a Γ hole band below $E_F$. As explained in Ref. 24, the relatively low dI/dV near $E_F$ is due to the M-centered electron bands (which dominate the DOS at $E_F$ here) having a shorter decay length into the vacuum compared to Γ-centered bands, resulting in much lower tunneling probability. The ARPES data of as-cleaved (Li$_{0.8}$Fe$_{0.2}$OH)FeSe, as presented in Fig. 1D, displays similar band structure as single-layer FeSe/SrTiO$_3$. Hence we would expect the resemblance in their tunneling spectra (on both FeSe surface). Below we refer to the Γ-centered electron-like band as the α band, Γ-centered hole-like bands as β bands and the M-centered electron-like band as the δ band.

**Evolution of the electronic states after K dosing**

Next, K atoms were deposited on the sample surface (see method section for details). Fig. 2 shows typical topographic images of the FeSe surface with $K_c$ from 0.008 ML to 0.306 ML. Here we define one monolayer (ML) as the areal density of Fe atoms in single-layer FeSe (1.41×10$^{15}$/cm$^2$). At small $K_c$, K atoms are randomly distributed on the surface (Fig. 2A-B). At certain coverages like 0.098 ML and 0.124 ML, K atoms can form locally-ordered structures, such as √5×√5 (with respect to the FeSe unit cell, Fig. 2C), or a six-fold close-packed lattice with an inter-atom spacing of 0.78 nm (Fig. 2D, see also Fig. S4A). There are different rotational domains observed in Fig. 2D (as marked by the arrows), due to different symmetry of the K lattice and underlying FeSe lattice. When $K_c > 0.15$ ML, K atoms begin to form clusters and no ordered surface structures can be observed (see Figs. S4C and S4D for larger scale images).

Fig. 3A and 3B show the detailed evolution of the dI/dV spectra as a function of $K_c$. At low coverage ($K_c < 0.080$ ML), it is seen from Fig. 3A that the onset of the α band gradually moves to lower energy. However, the β band does not shift together with α, instead moving slightly to higher energy. This anomalous behavior is possibly due to correlation effects in FeSe (20). In Fig. 3B one sees that the double superconducting gaps barely change for $K_c \leq 0.048$ ML. When $K_c$ reaches 0.062 ML ~ 0.075 ML, the bottom of the α band approaches $E_F$, thus the corresponding spectra in Fig. 3B tilt up at positive bias. However, the double coherence peaks at negative bias are still observable, which indicates the gap on δ band still exists. The corresponding gap size is only slightly suppressed ($\Delta_1$= 13.9 meV and $\Delta_2$= 8.6 meV for $K_c$ = 0.075 ML). This indicates that the superconductivity in the δ band is only weakly sensitive to additional electron doping.

When $K_c$ reaches 0.080 ML, the α band begins to cross $E_F$, as seen in both Figs. 3A and 3B. The tunneling conductance near $E_F$ is now greatly enhanced and dominated by the α band. The spectral weight from the δ band is overwhelmed and the double coherence peaks are no longer observable. (Note that the normalization scheme of Fig. 3B changes at this point to make all spectra appear with similar scale, see Fig. S5 for un-normalized dI/dV spectra near this Lifshitz transition). There is no gap-like feature near $E_F$ for $K_c$ = 0.080 ML or 0.098 ML, or the gap is much smaller than our experimental resolution (~1meV). This indicates the pairing is weak on the α band as it crosses $E_F$. In Fig. 4A, we summarize the energy shifts of the α and β bands as a function of $K_c$, by tracing the band bottom or top. We note that the sensitivity of the band position of α to surface K dosing is consistent with recent DFT calculations (25). It was shown that the $\alpha$ band has both Se 4$p$ and Fe 3$d$ orbital character, which makes it sensitive to Fe-Se distance or Se height ($h_{se}$) (24). K dosing could significantly affect the $h_{se}$ of the surface Se layer.

The Fermi surface of $\alpha$ will be a new electron pocket at $\Gamma$. To look for this pocket, we performed quasi-particle interference (QPI) mapping at $K_c = 0.124$ ML. As shown in Fig. 2D, for this coverage the K atoms form a close-packed structure with a relatively smooth, ordered surface, which is suitable for QPI measurements. The mapping was carried out in a 100×100 nm$^2$ area (Fig. 5A). Figs. 5B and 5C show a typical dI/dV map taken at $V_b = 10$ mV and its fast Fourier Transform (FFT), respectively. A complete set of dI/dV maps and FFTs taken within ±50 mV of $E_F$ can be found in Fig. S6. All FFTs display an isotropic scattering ring centered at $q = (0, 0)$, with the radius increasing with energy. In Fig. 5D we summarize the FFT linecuts though the center of the scattering ring, taken at various energies. An electron-like dispersion can be clearly seen, which is fully consistent with the presence of the $\alpha$ band. By assuming $q = 2k$ for the intra-band back-scattering condition, a parabolic fit yields the Fermi crossing at $k_F = 0.075$ Å$^{-1}$ and the band bottom at -37 meV (this value is also marked in Fig.4A). Such a sizable electron pocket has not been observed in iron-based superconductors at $\Gamma$ point before (for comparison, the $k_F$ of $\delta$ band for (Li$_{0.8}$Fe$_{0.2}$OH)FeSe is 0.21 Å$^{-1}$ at $K_c$=0, see Ref.16).

Shortly after the $\alpha$ band begins being occupied, starting from $K_c = 0.111$ ML, one sees a small gap open at $E_F$. We define the gap size by the peak or kinks on the gap edge, and refer it to $\Delta_3$ below. $\Delta_3$ reaches 3.5~4 meV for $K_c = 0.124$ ML, and closes at about $K_c = 0.136$ ML. In Fig. 5E we show a STS linecut taken on the surface in Fig. 2D ($K_c = 0.124$ ML) – the small gap is spatially uniform, with coherence peaks in most locations. We have checked this gap in several different samples and found that it can reach ~5meV at the "optimal" $K_c$ near 0.12 ML. Fig. 5F shows the temperature dependence of the gap at the optimal $K_c$, with clearly defined coherence peaks. It becomes less prominent as the temperature increases, vanishing at T = 35 K, close to the bulk $T_c$ of the sample (~42 K). Therefore, it is likely a possible superconducting gap opens on the $\alpha$ band, having a dome-like doping dependence. There could be other possibilities such as a charge-density-wave-induced gap; however, we did not observe any additional spatial modulation in the topographic image (Fig.2D, Fig.S4A), QPI maps (Fig.5, Fig. S6) and their FFTs (Fig. S4B). The gap has significant non-zero dI/dV at $V_b$=0, which could be due to gap anisotropy and/or thermal broadening effects. Measurements at lower temperature and high magnetic field would further clarify the nature of this gap.

The small gap disappears for $K_c = 0.136$ ML and 0.155ML, but starting from $K_c = 0.172$ ML, another gap-like feature develops at $E_F$. This time the gap size keeps increase upon further K dosing, and eventually at $K_c = 0.306$ ML, it exceeds 50 meV in width with a nearly flat bottom (Fig. 3B). We note that for $K_c = 0.201$ ML or 0.226ML, the gap has comparable size with the possible superconducting gap ($\Delta_3$) at $K_c$ =0.124 ML, but the feature is broader (bigger than $\Delta_3$ with weak or no coherence peak). Furthermore, at $K_c = 0.306$ ML, the gap is asymmetric with respect to $E_F$, and STM imaging is not possible for bias voltages inside the gap. Therefore, the gap opening starts from $K_c = 0.172$ ML likely evidences that the system enters an insulating state, with gradually depleted DOS at $E_F$. To illustrate this more quantitatively, In Fig.4B we integrated the dI/dV values extracted from Fig. 3A over the bias range of ±8 meV, as function of $K_c$ (> 0.1ML). This will give an estimation of the DOS of the α band near $E_F$ (note the integration window is larger than $\Delta_3$). It is clear that when $K_c$ < 0.172 ML, the DOS increases with $K_c$, while it quickly drops thereafter, indicative of a metal-insulator transition (MIT). This finding is consistent with the insulating state observed in K-dosed FeSe films by ARPES (20) and in ion-liquid gated (Li$_{1-x}$Fe$_x$OH)FeSe (26). Note that the topographic image of $K_c$ = 0.306 ML in Fig. 2F and Fig.S4D only show a disordered structure. This suggests the insulating phase is not due to the formation of some impurity phase (such as K$_2$Fe$_4$Se$_5$), but is intrinsic to deeply electron-doped FeSe. Moreover, the emergence of insulating phase also indicates K atoms do not form a surface metallic layer by themselves up to $K_c$ = 0.306 ML. The STS in Fig. 3 shall reflect the electron states of doped FeSe layer.

To facilitate the understanding of the STM data, we performed AREPS measurements on K-dosed (Li$_{0.8}$Fe$_{0.2}$OH)FeSe (experiment details is described in Method section). Fig. 6A and 6B show ARPES intensity along the cuts crossing Γ and M (Fig. 6C), respectively, as the function of K coverage ($K_c$). Note that the $K_c$ here is estimated from K flux and deposition time ($t$) (see Method section). As seen from Fig. 6B, the size of the δ Fermi pocket increases with K dosing (for $K_c$≤~0.27ML), indicating the electron doping. Meanwhile, near Γ point (Fig. 6A), there is noticeable spectral intensity shows up and increases near $E_F$ upon K dosing (for $K_c$<~0.27ML). To illustrate it more quantitatively, we plot the corresponding MDC and EDC curves (taken near $E_F$ and k=0) for various $K_c$ in Figs. 6D and 6E, respectively (see figure captions). The spectral intensity at Γ evidences the emergence of an electron pocket, though the band dispersion is not clear which could be due to small pocket size and/or limited resolution here. To have a comparison with STM result, in the $K_c$ ~0.12 ML panel of Fig. 6A we superposed the band dispersion of α which is derived from the QPI of $K_c$ =0.124 ML (Fig. 5D). There is a qualitative match between QPI band dispersion and ARPES intensity at Γ. Furthermore, it is noticeable that at high dosing ($K_c$~0.45ML, $t$=302s), the bands at both Γ and M near $E_F$ became unresolvable, which is also consistent with a metal-insulator transition suggested by the STM data. In Fig. 6F we show symmetrized EDC taken near the $k_F$ of δ band (marked in Fig. 6B), which displays the evolution of the superconducting gap on δ band. The gap size was ~13 meV for $K_c$=0 and $K_c$~0.06ML, decreased to ~9meV for $K_c$~0.12ML and disappeared for $K_c$~0.27ML. The disappearance of superconductivity on δ band before entering insulating phase is also observed in K-dosed FeSe films (20).

We noted that the ARPES signal should come from both FeSe and Li$_{0.8}$Fe$_{0.2}$OH surface (the light spot is of millimeter size here). Our previous STM study found a small electron pocket at Γ for Li$_{0.8}$Fe$_{0.2}$OH surface (16), and it may account for the weak spectral weight at Γ near $E_F$ for the $K_c$=0 case in Fig. 6A (also indicated in Figs. 6D and 6E). We note that a recent μSR study reported proximity-induced superconducting gap in the Li$_{1-x}$Fe$_x$OH layers, which also suggest the Li$_{1-x}$Fe$_x$OH layer is conductive (27).

Fig. 7 summarizes the observed electronic states from the STS in Fig. 3, as a function of $K_c$. This phenomenological phase diagram contains four distinct regimes. In the first (0≤$K_c$≤ 0.075 ML), the Fermi surface only comprises M-centered δ band, and its superconducting gap ($\Delta_1$ and $\Delta_2$) is only gradually suppressed. In Regime II (0.080 ML≤ $K_c$≤ 0.172 ML), the α band crosses $E_F$, introducing a new electron pocket at Γ (illustrated in the inset). A possible new superconducting dome on $\alpha$ band exists in the middle of this regime (Green squares represent the gaps size of $\Delta_3$). As a complement, the ARPES measured gap sizes on $\delta$ band (from Fig. 6F) are also marked here by gray circles. It appears the gap persists in the left part of Regime II, thus STM measured $\Delta_1$ and $\Delta_2$ should also extend to Regime II (indicated by two short dashed lines). In Regime III (0.172 ML < $K_c$ ≤ 0.26 ML), the DOS near $E_F$ begins to decrease as the system approaches a metal-insulator transition. Finally, in Regime IV ($K_c$ > 0.26 ML), the DOS near $E_F$ is depleted and the system enters an insulating state.

We noted that the Fermi surface of $A_xFe_{2-y}Se_2$ at $k_z = \pi$ plane (10) is similar to the one shown in Regime II of Fig. 7. However, the center electron pocket does not exist at $\Gamma$ ($k_z = 0$) in $A_xFe_{2-y}Se_2$, reflecting its significant 3D character. In $(Li_{0.8}Fe_{0.2}OH)FeSe$, the interlayer spacing between two FeSe layers (~0.932nm, Ref. 4) is significantly larger than that of $A_xFe_{2-y}Se_2$ (~0.702nm, Ref. 1). This makes the Fermi surface of $(Li_{0.8}Fe_{0.2}OH)FeSe$ rather two dimensional (14).

**Discussion**

Surface K-dosed $(Li_{0.8}Fe_{0.2}OH)FeSe$ provides several unique opportunities to understand the superconductivity in Fe-based superconductors. Firstly, the emergence of the $\Gamma$-centered electron pocket will introduce a new pairing channel. For most known iron-based superconductors, there are two typical types of Fermi surface topology: one with hole pockets at the zone center and electron pockets at the zone corner, the other with only electron pockets at the zone corner. The scattering between different Fermi pockets has direct consequences on the pairing symmetry (28-31). It was suggested that the interband interactions (spin fluctuations) between the $\Gamma$-hole and M-electron pockets with wave vector Q= ($\pi$, 0) are the main pairing glue, which will lead to $s_\pm$ wave pairing symmetry (28,29). However, the absence of a $\Gamma$ pocket in electron-doped FeSe-based systems seriously challenges this scenario. Later, it was suggested that the interaction between neighboring M-electron pockets with Q= ($\pi$, $\pi$) would dominate pairing in such cases and lead to a $d$-wave pairing symmetry (29-31), but this picture lacks direct experimental support. Recently, some theoretical work shows that the "incipient" band (a band which is close to but does not cross $E_F$) may still play an important role in pairing, with a significant pairing potential (32~34), and a large "shadow gap" feature was indeed observed in the incipient $\Gamma$ band in $LiFe_{1-x}Co_xAs$ (35). Here, by surface K-dosing $(Li_{0.8}Fe_{0.2}OH)FeSe$, we are able to continuously tune the $\alpha$ band to approach and cross $E_F$, which is expected to enable the interaction between two electron bands at $\Gamma$ and M with Q near ($\pi$, 0). (For $A_xFe_{2-y}Se_2$, such interactions may exist but would be weaken by the strong 3D character of its central electron pocket, as aforementioned). We did not observe gap opening on the $\alpha$ band near its Lifshitz transition (0.062ML ≤ $K_c$ ≤ 0.098 ML), while the gap on the $\delta$ band is slightly suppressed. This would suggest such a $\Gamma$-M interaction does *not* promote superconductivity at the onset of the transition, and the dominant pairing interaction must still lie in the $\delta$ band. When the $\alpha$ band does develop a gap in Regime II, assuming the observed gap is possibly a superconducting gap, the small gap size (compared to that on the $\delta$ band) also suggests a weak pairing potential on the $\alpha$ band. In fact, since the gap-closing temperature is quite high, this gap could be induced by the $\delta$ band through normal interband scattering, since the latter band remains superconducting as indicated in Fig. 6F and Fig. 7. Nevertheless, the dome-like behavior suggests that the $\alpha$ band gradually participates in the pairing. Due to the close competition of various pairing channels, the new type of Fermi surface topology found here may help facilitate a novel superconducting pairing state. In addition, orbital-selective pairing (36,37), as recently evidenced in bulk FeSe (38), may also relate to our results. Band calculation of single-layer FeSe shows the major orbital component of $\alpha$ is $d_{x2-y2}$ (24), which differs from the $d_{xy}$ and $d_{xz}/d_{yz}$ orbitals that composed $\delta$ band (29). Further theoretical work considering all possible inter-, intra-band interactions and orbital structures will be needed to understand the electron pairing in such a case.

Secondly, the metal-insulator transition observed here provides more clues as to the unusual doping-driven insulating phase in FeSe. In particular, our result shows that the DOS near $E_F$ is gradually depleted during the transition, over a relatively wide doping range (from $K_c = 0.172$ ML to ~0.26 ML). This differs from transport measurements in ionic liquid-gated $(Li_{1-x}Fe_xOH)FeSe$, where a sharp, first-order-like transition is observed (26). The smooth transition is consistent with the ARPES result on K-dosed FeSe, where a gradual suppression of spectral weight accompanied by an increasing effective mass is observed (20), suggestive of a correlation-driven transition (39). We note a similar insulating phase has been observed in $Rb_xFe_{2-y}Se_{2-z}Te_z$ (40), which indicates the correlation-driven metal-insulator transition might be universal in FeSe-derived superconductors.

Thirdly, K dosing may be able to change the band topology of the top FeSe layer, inducing a topological phase transition. Recently Wu *et al.* proposed that the band topology of the Fe(Te)Se system is controlled by Se(Te) height, which affects the separation ($\Delta_n$) between the electron and hole bands at $\Gamma$ (41), and suggested that if $\Delta_n$ is smaller than 80 meV, spin-orbit coupling can induce band inversion and lead to a nontrivial $Z_2$ topology. In our case, the separation between the $\alpha$ and $\beta$ bands is continuously reduced from 120meV ($K_c = 0$) to ~20meV ($K_c$ ~0.1ML), as summarized in Fig. 4A. Therefore, such a topological phase transition may well be achievable. We noted that for $K_c >$0.1ML the evolution of $\alpha$ and $\beta$ bands are hard to identify in STS (Fig. 3A), however topological edge states may exist near step edges if the system enters a nontrivial phase, which deserves further investigation.

In summary, by dosing K on the surface of $(Li_{0.8}Fe_{0.2}OH)FeSe$, a new electron pocket can be introduced at the $\Gamma$ point. This Lifshitz transition creates a new type of Fermi surface topology and enables a new pairing channel *via* $\Gamma$-M interactions. However, only a small gap feature was observed on the new $\Gamma$ pocket, indicating its weak pairing potential. Further doping eventually drives the system into an anomalous insulating state. In addition, nontrivial band topology might be realized by the K-dosing-induced band shift. This singular combination of new opportunities makes K-dosed $(Li_{0.8}Fe_{0.2}OH)FeSe$ an intriguing platform for studying the pairing interaction, correlation effects and topological properties in iron-based superconductors.

Upon completing this work, we noticed an ARPES study on surface K-dosed 1UC FeSe/SrTiO$_3$ (42), which has similar band structure to $(Li_{0.8}Fe_{0.2}OH)FeSe$. An electron pocket at $\Gamma$ is also observed after K dosing. This suggests the broader applicability of our findings.

**Materials and Methods**

**Sample growth:**

($Li_{0.8}Fe_{0.2}$OH)FeSe single crystals were grown by hydrothermal ion-exchange method described in Ref. 21. $K_{0.8}Fe_{1.6}Se_2$ matrix crystal, $LiOH·H_2O$, Fe, and $CH_4N_2Se$ were used as starting material. During the hydrothermal reaction, $Li_{1-x}Fe_xOH$ layers were formed and replaced the K atoms in $K_{0.8}Fe_{1.6}Se_2$ (21). Resistivity and magnetic susceptibility measurements (Fig. S1A and B) confirm the $T_c$ of about 42K. The optical image (Fig. S1C) shows that the sample surface is composed of separated domains with the size of tens of microns. Such morphology may be due to the ion-exchange process.

**STM measurement:**

STM experiment was conducted in a commercial Createc STM at the temperature of 4.5K. ($Li_{0.8}Fe_{0.2}$OH)FeSe samples were cleaved in ultrahigh vacuum at 78K. Pt tips were used in all measurements after careful treatment on a Au(111) surface. The tunneling spectroscopy (dI/dV) was performed using a standard lock-in technique with modulation frequency $f = 915$ Hz and typical amplitude $\Delta V = 1$ mV.

**ARPES measurement:**

ARPES measurement was conducted in an in-house ARPES system with a Helium discharged lamp (21.2 eV photons), at the temperature of 11K, using Scienta R4000 electron analyzers. The energy resolution was 8 meV, and the angular resolution was 0.3 degrees. ($Li_{0.8}Fe_{0.2}$OH)FeSe samples were cleaved in-situ under ultrahigh vacuum. During measurements, the spectroscopy qualities were carefully monitored to avoid the sample aging issue.

**K dosing:**

K atoms were evaporated from standard SAES alkali metal dispenser, and the samples are keep at 80K during K dosing. In STM study, the $K_c$ at low coverages is obtained by directly counting surface K atoms. Then the K deposition rate is carefully calibrated and the $K_c$ at high coverage is calculated by deposition rate and time. The $K_c$ dependence of the STS is obtained by repeated deposition of K atoms on one sample. After each deposition, the STM tip is placed nearly on the same surface domain which is found to be mostly covered by FeSe terminated surface. In ARPES study, $K_c$ is estimated from K flux rate (measured by a quartz crystal microbalance) and deposition time. $K_c$ dependence of the ARPES spectra are obtained by repeated deposition of K atoms on one sample.

**Acknowledgments:** We thank Dr. D. C. Peets and Dr. B. Y. Pan for helpful discussions.
**Funding:** This work is supported by the National Science Foundation of China and National Key R&D Program of the MOST of China (Grant No. 2016YFA0300200).
**Author contributions:** M. Q. Ren, Y. J. Yan and R. Tao performed the STM/STS measurement and analyzed the data. X. H. Niu and R. Peng performed the ARPES measurement and analyzed the data. D. Hu synthesized the sample under the guidance of J. Zhao. T. Zhang and D. L. Feng designed and coordinated the whole work and wrote the manuscript. All authors have discussed the results and the interpretation.
**Competing interests:** The authors declare that they have no competing interests.
**Data and materials availability:** All data needed to evaluate the conclusions in the paper are present in the paper and/or the Supplementary Materials. Additional data available from authors upon request.

**Figures**

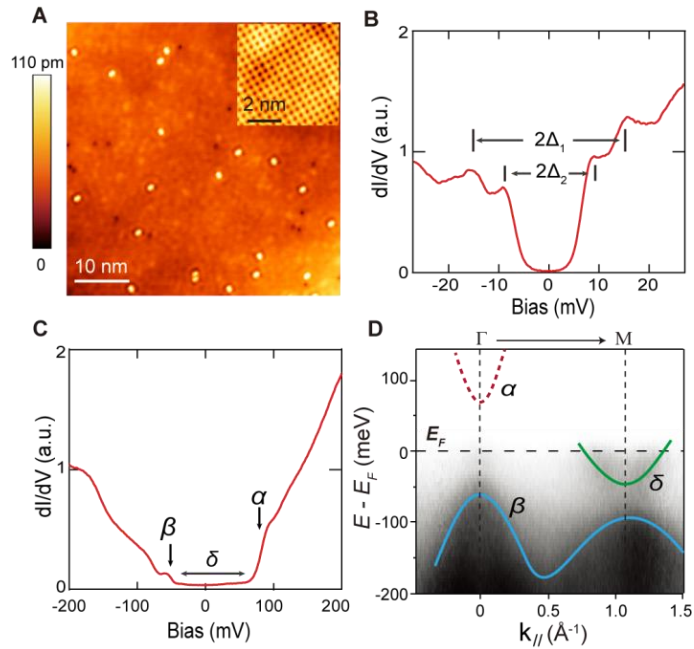

**Fig. 1. Topographic image, tunneling and ARPES spectra of as-cleaved (Li$_{0.8}$Fe$_{0.2}$OH)FeSe.** (**A**) Topographic image of as-cleaved, FeSe-terminated surface ($V_b$=100 mV, $I$=50 pA), inset shows the surface lattice. (**B**) Low-energy dI/dV spectrum of as-cleaved FeSe surface, which displays double superconducting gaps of size $\Delta_1$=15meV and $\Delta_2$=9meV. (**C**) Larger energy scale dI/dV spectrum. Arrows indicate the onset of the α and β bands (see text). Horizontal bar indicates the range of δ band. (**D**) ARPES measurement of as-cleaved (Li$_{0.8}$Fe$_{0.2}$OH)FeSe. Solid curves track the dispersion of the β and δ bands, while the α band above $E_F$ is sketched with red dashed curve.

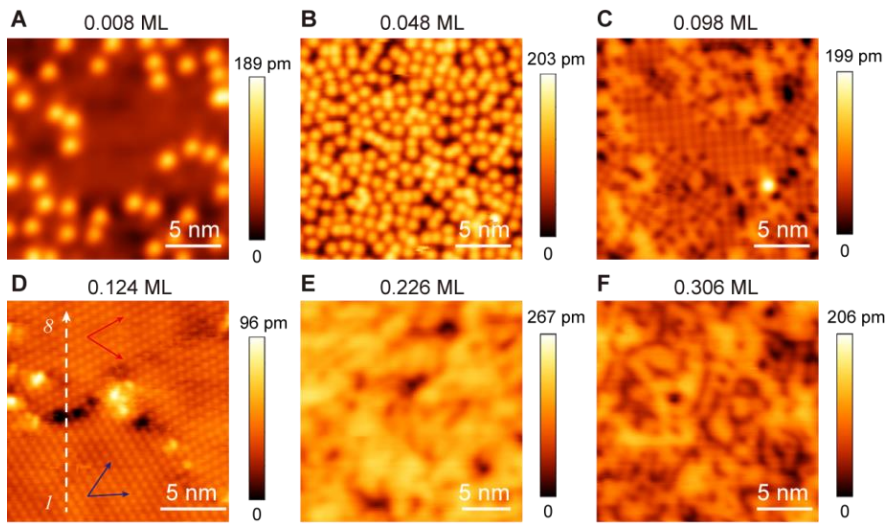

**Fig. 2. Topographic images of FeSe surface with different K coverage ($K_c$).** (**A**) $K_c$ = 0.008ML, (**B**) $K_c$ = 0.048 ML, (**C**) $K_c$ = 0.098 ML, (**D**) $K_c$ = 0.124 ML, (**E**) $K_c$ = 0.226 ML, (**F**) $K_c$ = 0.306 ML. Typical imaging parameters are $V_b$ = 0.5 V and $I$ = 50 pA. The red and blue arrows in (D) indicate the orientation of two different rotational domains. The white dashed arrow marks the position where the STS in Fig. 5E are taken.

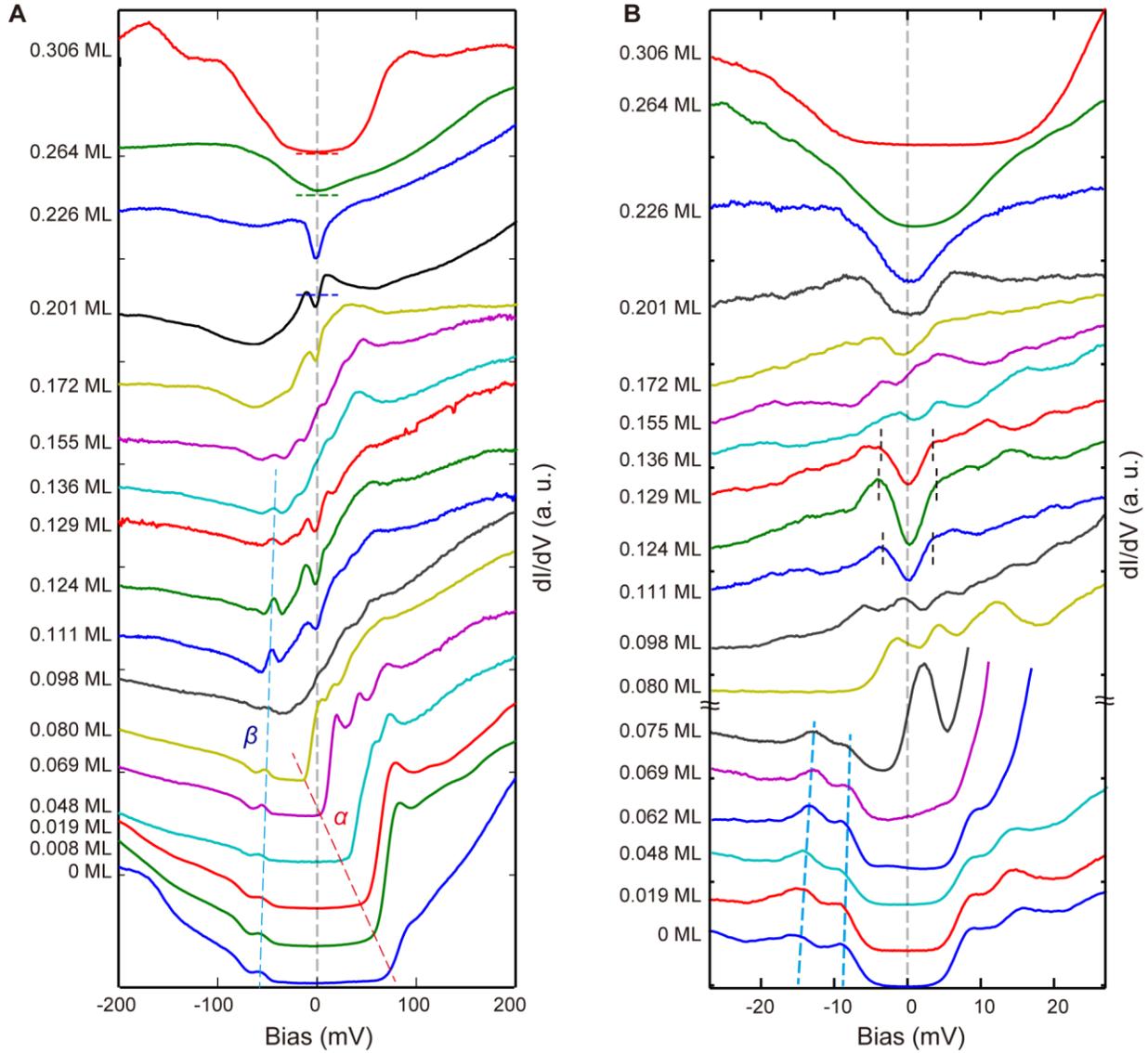

**Fig. 3. Evolution of dI/dV spectra taken on FeSe surface with various $K_c$ as labeled.** (**A**) Typical dI/dV spectra taken within large energy range (±200 meV). Red and blue dashed lines track the onsets of the α and β bands. The zero positions of the spectra for $K_c$ = 0.306 ML, 0.264 ML, 0.226 ML are marked by short horizontal bars. (**B**) Typical dI/dV spectra taken near $E_F$ (±27meV). Two blue dashed lines track the superconducting coherence peaks at negative bias. The curves for $K_c$ ≤0.075 ML are normalized by the dI/dV value at $V_b$ = -27 mV, and curves for $K_c$ >0.075 ML are normalized by the value at $V_b$ = 27 mV. $E_F$ ($V_b$ =0) is indicated by gray dashed lines. For $K_c$ = 0.111 ML, 0.124ML and 0.129ML, the gap edge positions (define $\Delta_3$) are marked by short dashed lines.

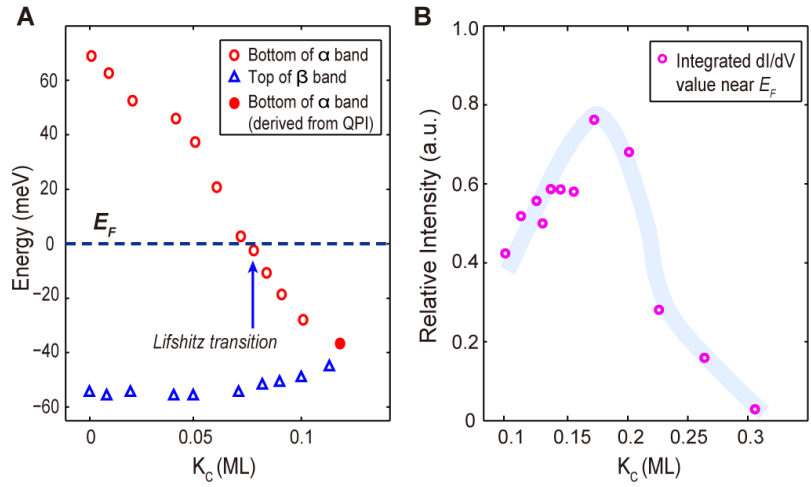

**FIG. 4** (**A**) The doping dependence of the band bottom (top) energy of the $\alpha$ ($\beta$) band. At $K_c =$ 0.080 ML, the $\alpha$ band begins to cross $E_F$. (**B**) Integrated dI/dV values within the bias range of ±8 meV as function of $K_c$, which reflects the DOS near $E_F$.

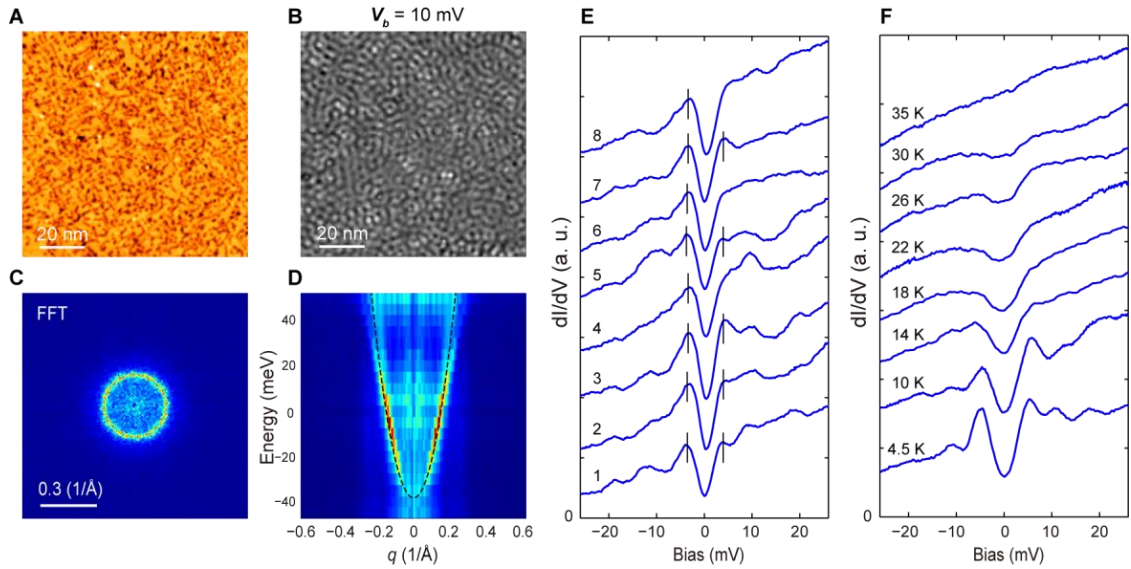

**FIG. 5. QPI measurement of the $\alpha$ band and the spatial and temperature dependence of its gap.** (**A**) Topographic image of the mapping area of size 100×100 nm² ($K_c$ = 0.124 ML). (**B**) Typical dI/dV map taken at $V_b$ = 10 mV. The setpoint for dI/dV map is: $V_b$=50mV, $I$=150pA and $\Delta V$ = 3mV. (**C**) FFT image of (B). (**D**) Intensity plot of the FFT line cuts though $q = (0, 0)$, dashed curve is the parabolic fit. Note that the small gap is not observable here because of the large modulation ($\Delta V$). (**E**) A dI/dV line cut taken along the dashed arrow in Fig. 2D, showing a spatially uniform gap. Bars indicate the coherence peaks. (**F**) Temperature dependence of the gap taken on a different sample with $K_c$ ~0.12 ML.

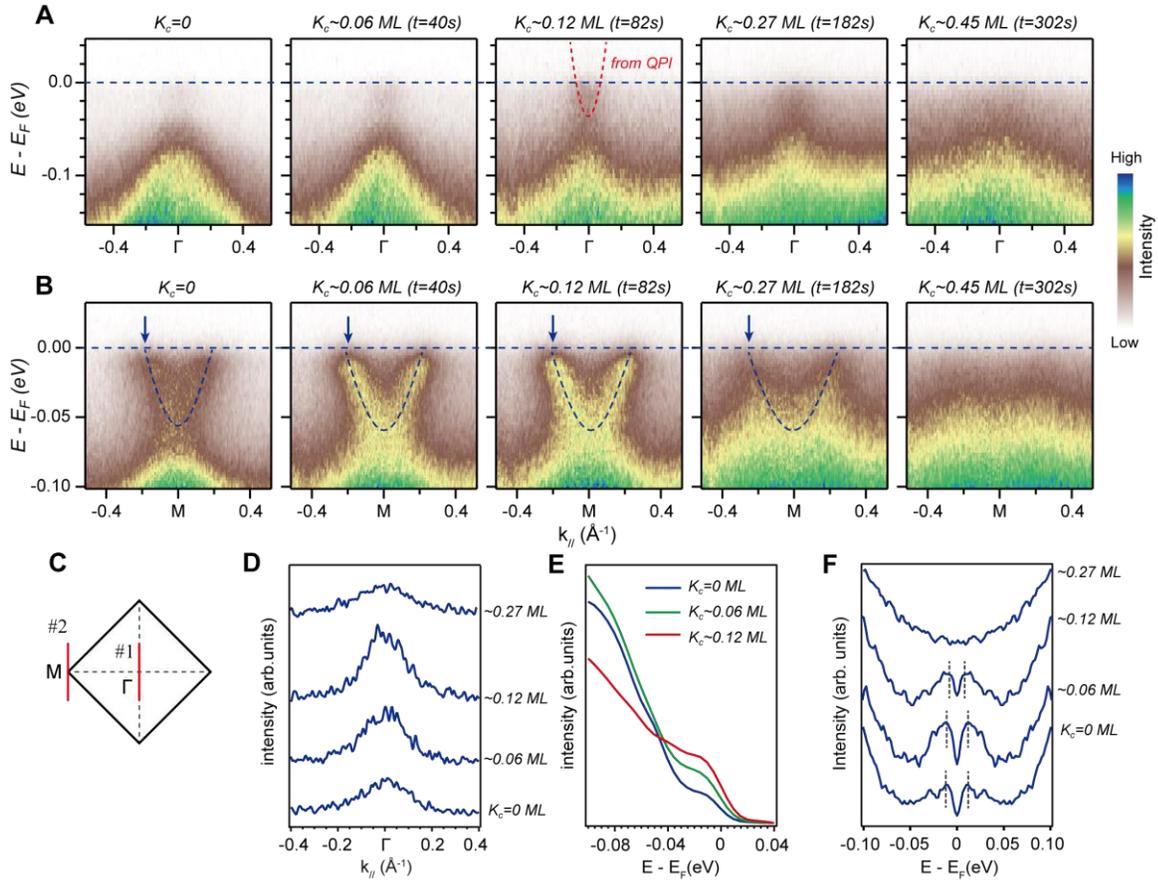

**FIG. 6. ARPES measurement of the band structure of surface K-dosed (Li$_{0.8}$Fe$_{0.2}$OH)FeSe.** (**A**) ARPES intensity along the cut #1 shown in (C), as a function of $K_c$ and deposition time (*t*). Red dashed line in the third panel ($K_c$ ~0.12 ML) represents the band dispersion of α that derived from QPI (Fig. 5D). (**B**) ARPES intensity along the cut #2 shown in (C), as a function of $K_c$ and *t*. Dashed lines track the dispersion of δ band. (**C**) Sketch of the Brillouin zone of (Li$_{0.8}$Fe$_{0.2}$OH)FeSe. (**D**) Evolution of the momentum distribution curve (MDC) along cut #1 upon K dosing, integrated over ±14 meV at $E_F$ (curves are shifted vertically for clarify). The intensity at Γ increases up to $K_c$~0.12 ML. The decreased intensity at $K_c$ ~0.27 ML could be due to the approaching to the insulating phase (consistent with Fig. 4B). (**E**) Evolution of the energy distribution curve (EDC) taken around k=0 (Γ point) upon K dosing ($K_c$ =0 ~ 0.12 ML). The increased intensity between -0.04eV ~ 0eV is consistent with the emergence of an electron pocket. (**F**) Symmetrized EDC showing the evolution of the superconducting gap on the δ band, as a function of $K_c$. The momenta of individual spectra are indicated by the arrows in (B).

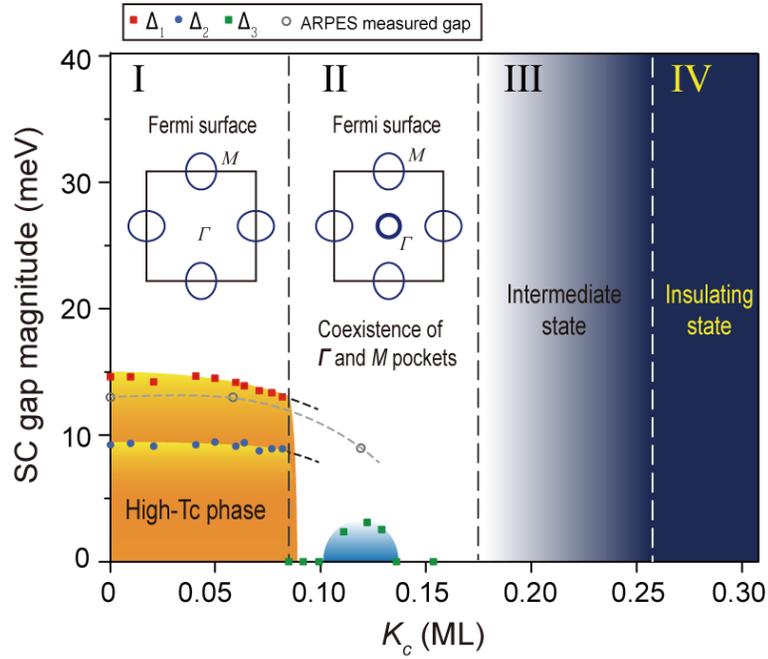

**FIG. 7. Summarized phase diagram of surface K-dosed (Li$_{0.8}$Fe$_{0.2}$OH)FeSe.** The insets in regimes I and II sketch the Fermi surface before and after the Lifshitz transition. The red, blue and green dots represent the value of $\Delta_1$, $\Delta_2$, $\Delta_3$, respectively. Gray circles represent the ARPES measured gap size on $\delta$ band (gray dashed line traces its variation). ARPES measurement suggests $\Delta_1$ and $\Delta_2$ would not suddenly disappear when entering regime II, as illustrated by the short black dashed lines.

# Supplementary Materials for "Superconductivity across Lifshitz transition and anomalous insulating state in surface K-dosed (Li$_{0.8}$Fe$_{0.2}$OH)FeSe"

M. Q. Ren[1], Y. J. Yan[1], X. H. Niu[1], R. Tao[1], D. Hu[1], R. Peng[1], B. P. Xie[1,2], J. Zhao[1,2], T. Zhang[1,2*], D. L. Feng[1,2*]

[1]State Key Laboratory of Surface Physics, Department of Physics, and Advanced Materials Laboratory, Fudan University, Shanghai 200433, China
[2]Collaborative Innovation Center of Advanced Microstructures, Nanjing, 210093, China

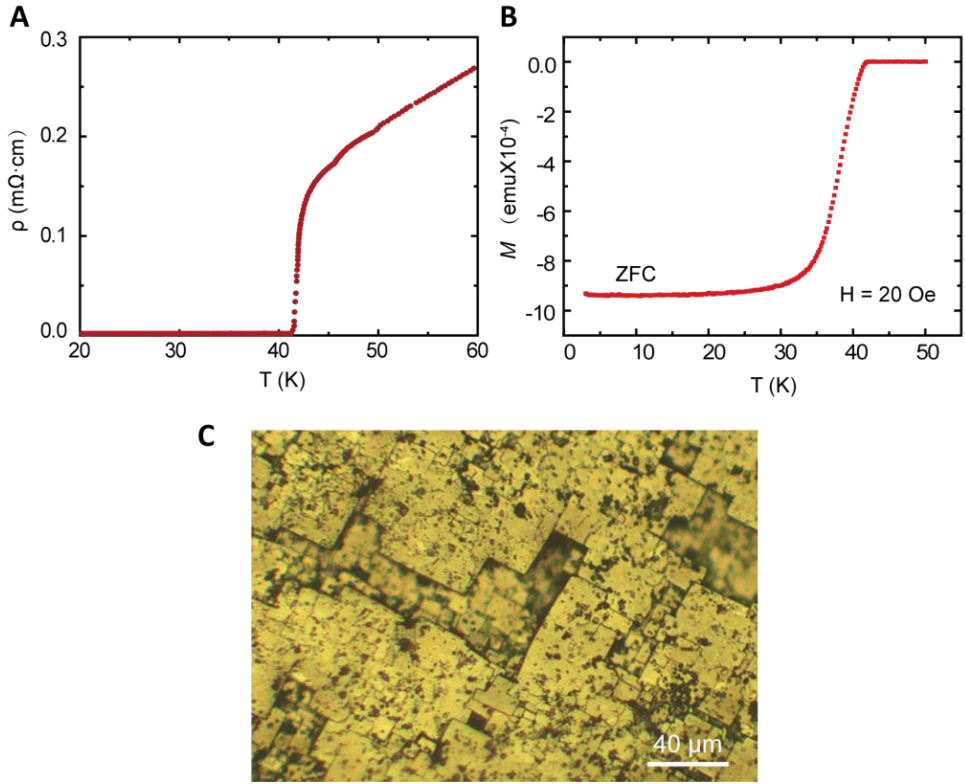

**Fig. S1.** (**A**) Temperature dependence of the resistivity of (Li$_{0.8}$Fe$_{0.2}$)OHFeSe single crystal. (**B**) Temperature dependence of the DC magnetic susceptibility of (Li$_{0.8}$Fe$_{0.2}$)OHFeSe measured with zero-field cooling (ZFC). (**C**) Optical microscopy image of a surface of (Li$_{0.8}$Fe$_{0.2}$)OHFeSe single crystal.

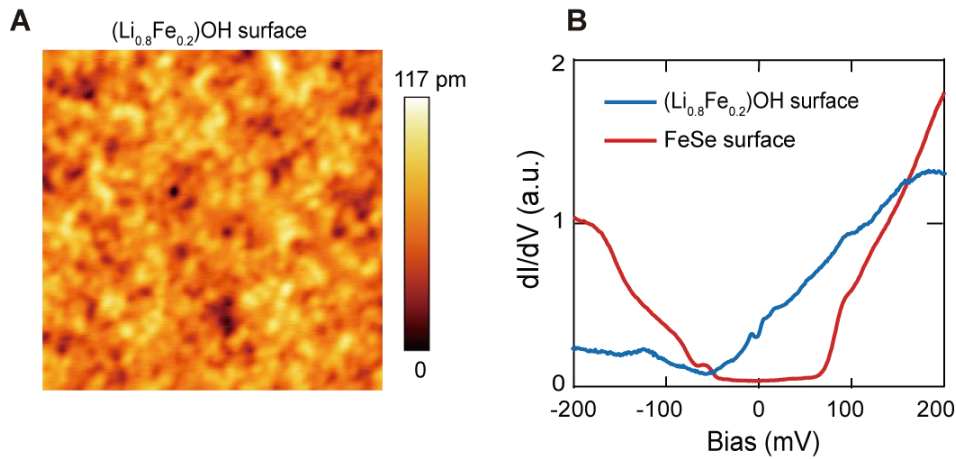

**Fig. S2.** (**A**) Topographic image of as-cleaved Li$_{0.8}$Fe$_{0.2}$OH surface. (**B**) dI/dV spectra taken on Li$_{0.8}$Fe$_{0.2}$OH surface and FeSe surface, respectively.

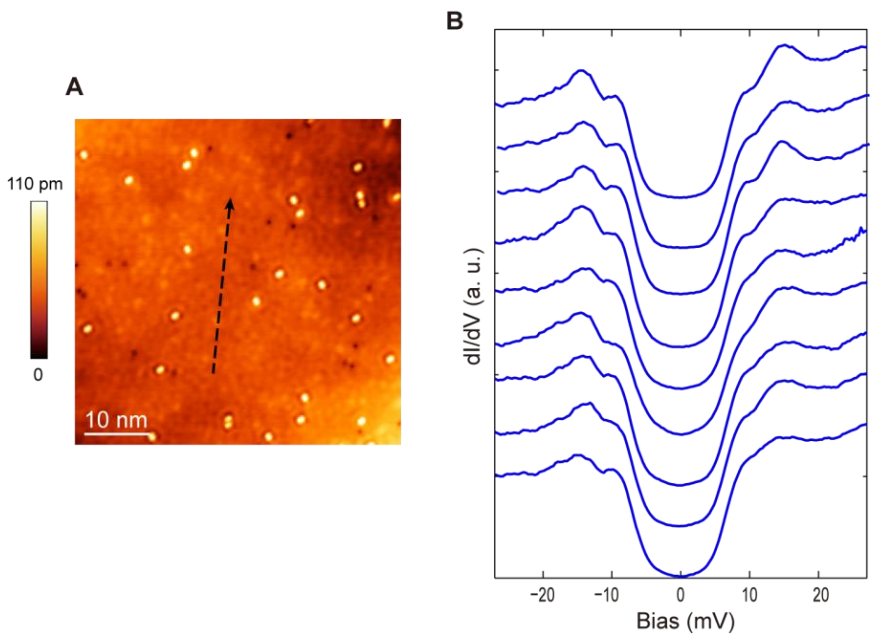

**Fig. S3. Spatial distribution of the superconducting gap on as-cleaved FeSe surface.** (**A**) Topography of FeSe surface (same as Fig.1A). (**B**) dI/dV spectra taken along the line cut marked in (A) shows a spatially-homogenous superconducting gap.

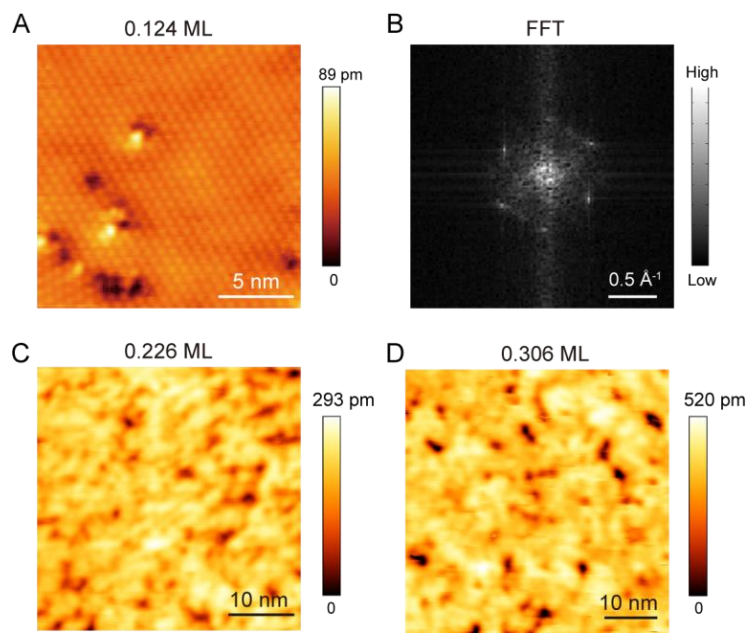

**Fig. S4. Additional topographic images of FeSe surface after K dosing. (A).** $K_c$=0.124 ML, taken mostly in a single rotational domain. **(B)** FFT image of (A), showing six Bragg spots. (Note that due to the tip drift in scanning the Bragg spots are not perfectly six-folding symmetric). **(C)** $K_c$=0.226 ML (size: 40×40 nm$^2$), **(D)** $K_c$=0.306 ML (size: 50×50 nm$^2$)

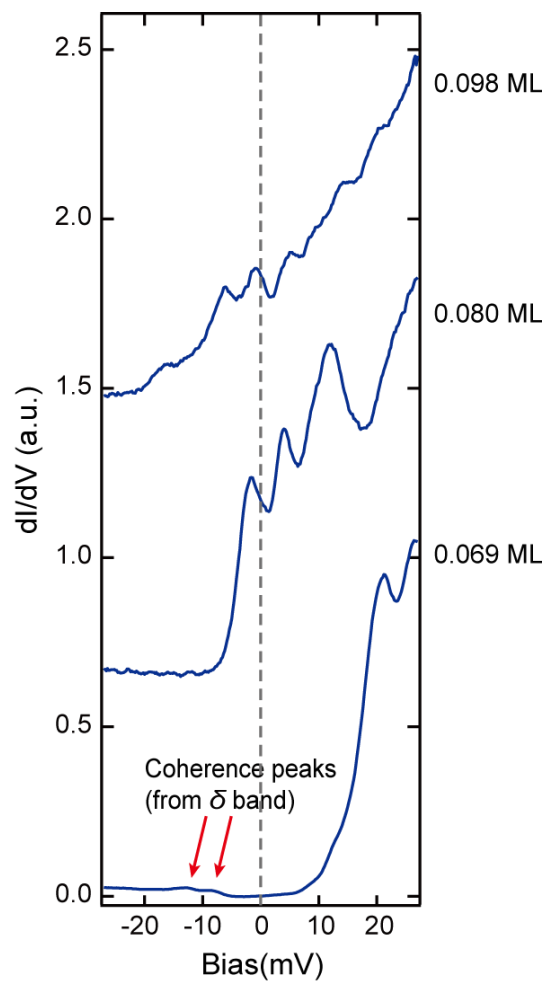

**Fig. S5.** Un-normalized dI/dV spectra of $K_c$=0.069 ML, 0.080ML and 0.098 ML, showing the evolution of the DOS near the Lifshitz transition. The red arrows indicate the double coherence peaks of the δ band.

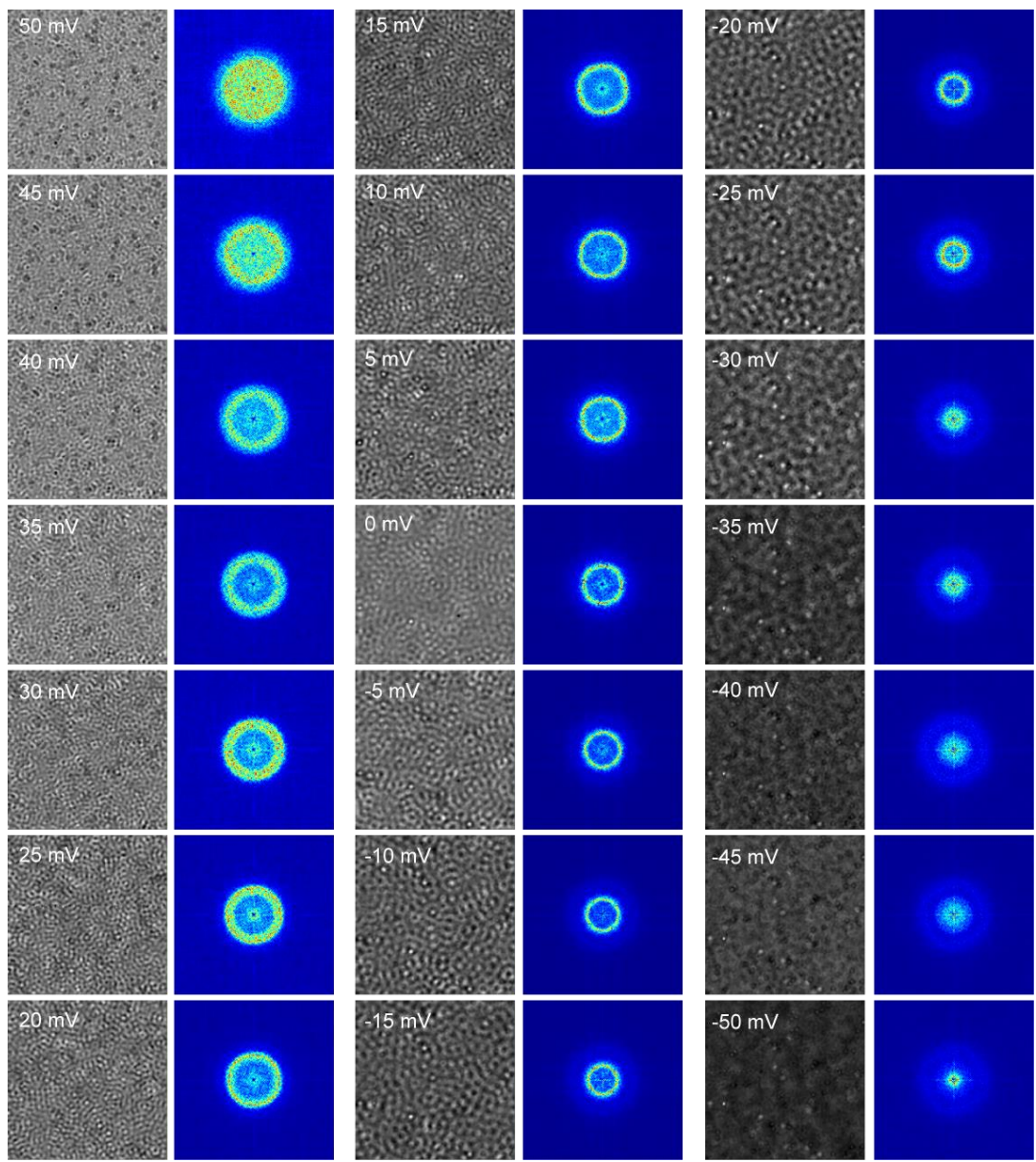

**Fig. S6.** dI/dV maps and corresponding FFTs taken in a 100×100 nm² area of the FeSe-terminated surface, with $K_c$=0.124 ML. Set point: $V_b$ = 50 mV, $I$ = 150 pA, $\Delta V$ = 3 mV. Each map has 200 × 200 pixels. The FFT images are four-fold symmetrized.